\documentclass[12pt]{article}
\usepackage{amsmath,amsfonts,amssymb}
\usepackage{cases}

\usepackage{indentfirst}
\makeatletter
\renewcommand{\@begintheorem}[2]{\begin{trivlist}
\item[\hspace{\labelsep}{\bfseries#1\ #2.}]\itshape}
\renewcommand{\@opargbegintheorem}[3]{\begin{trivlist}
\item[\hspace{\labelsep}{\bfseries#1\ #2\ (#3).}]\itshape}
\renewcommand{\@endtheorem}{\end{trivlist}}
\makeatother
\author{Sergey V.\,Smirnov\thanks{Department of Mathematics and Mechanics, Moscow State University. E-mail: {\tt ssmirnov@higeom.math.msu.su}}}
\title{Factorization of Darboux--Laplace transformations for discrete hyperbolic operators}

\def\pa{\partial}

\def\phi{\varphi}

\newcommand{\ord}{\mathop{\rm ord}\nolimits}
\newcommand{\Ker}{\mathop{\rm Ker}\nolimits}

\newcommand{\mb}[1]{{\textbf {\textit#1}}}

\newtheorem{theorem}{\sc Theorem}
\newtheorem{lemma}{\sc Lemma}
\newtheorem{remark}{\sc Remark}

\newtheorem{definition}{\sc Definition}

\begin{document}

\maketitle
\begin{abstract}
Elementary Darboux--Laplace transformations for semidiscrete and discrete second order hyperbolic operators are classified. It is proved that in the (semi)-discrete case there are two types of elementary Darboux--Laplace transformations as well: Darboux transformations that are defined by the choice of particular element in the kernel of the initial hyperbolic operator and classical Laplace transformations that are defined by the operator itself. It is proved that in the discrete case on the level of equivalence classes any Darboux--Laplace transformation is a product of elementary ones.
\end{abstract}

\section{Introduction}

In general, {\it Darboux transformation} is a mapping ${\cal L}\mapsto\hat{\cal L}$ between two differential or difference operators defined by operator relation
\begin{equation}
\label{dlt}
\hat{\cal L}\circ{\cal D}=\hat{\cal D}\circ{\cal L},
\end{equation}
where ${\cal D}$ and $\hat{\cal D}$ are some differential (difference) operators. Darboux transformations play an important role in theory of integrable systems~\cite{MS1}. They can be used for finding explicit solutions of integrable equations~\cite{Ma,MS2}, for obtaining Lax representations~\cite{VSh,Ha,Sm1} or for finding characteristic integrals for hyperbolic equations~\cite{Sm2}. Darboux transformations serve as discrete symmetries of integrable equations in~\cite{VSh}. They were also widely used by G.~Darboux in his classical book on theory of surfaces~\cite{Da}.

In the one-dimensional case relation~(\ref{dlt}) yields the condition $\hat{\cal D}={\cal D}$ for Schr\"odinger operators. Operator relation
\begin{equation}
\label{dt}
\hat{\cal L}\circ{\cal D}={\cal D}\circ{\cal L}.
\end{equation}
for Schr\"odinger operator on the line has been studied in~\cite{VSh,Sh1,BS}.

First-order Darboux transformations (i.e. operator relations~(\ref{dlt}) or~(\ref{dt}) with ${\cal D}$ and $\hat{\cal D}$ being first-order operators) are called {\it elementary}. Iteration of elementary transformations satisfying relation~(\ref{dt}) in the one-dimensional case leads to the so-called {\it Veselov--Shabat dressing chain} that is an integrable hamiltonian system~\cite{VSh}. Iteration of classical {\it Laplace transformations} satisfying relation~(\ref{dlt}) in the two-dimensional case is known to be described by {\it two-dimensional Toda lattice} (see~\cite{Sh2}); if one imposes special boundary conditions, this system of PDEs is Darboux-integrable~\cite{Le}. Iteration of Laplace transformations in the discrete case leads to {\it discrete Toda lattice}~\cite{AS}. Therefore, it is interesting to understand whether, in general, Darboux transformations are factorizable into a product of elementary ones or not.

In the one-dimensional case each elementary Darboux transformation is defined by a certain particular element from the kernel of operator ${\cal L}$, and all Darboux transformations are factorizable into a product of elementary ones for differential operator ${\cal L}$ of arbitrary order. This fact is well-known to specialists, and simple proof can be found in~\cite{AMSh}. Exactly in the same way factorization theorem can be proved for one-dimensional difference operators. In the two-dimensional case the situation is much more complicated: if ${\cal L}$ is a hyperbolic second-order differential operator, then there are two essentially different classes of elementary Darboux transformations: transformations depending on some particular solution and classical Laplace transformations that depend only on initial operator itself. Both types of these transformations were already known to Darboux~\cite{Da}, but rigorous proof of the fact that the list of elementary transformations is exhausted by these two classes have probably appeared in literature for the first time only less then a decade ago in~\cite{She1} (this result was known to some specialists before, but only as an example of ``mathematical folklore'').

All Darboux transformations of total order two (that is, the sum of the orders in two variables) for two-dimensional hyperbolic operators have been classified in~\cite{She2}. Factorization theorem for Darboux transformations of special kind for two-dimensional hyperbolic operators with vanishing Laplace invariants has been proved in~\cite{She3}. General factorization result on Darboux transfromations for two-dimensional hyperbolic operators has been obtained in~\cite{She4}. Darboux transformations for operators of order higher than two in the two-dimensional case were discussed in~\cite{She5} and transformations for multidimensional differential operators were considered in~\cite{HSh}. Factorization theorem for Darboux transformations for differential operators of arbitrary order on the superline was proved in recent papers~\cite{HShV,LShV}.

In the discrete case general theory of Darboux transformations had not been discussed much up to now. Discrete Darboux transformations (or Darboux--Laplace transformations) in the two-dimensional case were applied in~\cite{Sm1,Sm2,DN}. Discrete {\it wronskian} (or {\it casoratian}) formulas were used in~\cite{DT}. A $q$-version of discrete Darboux transformation in the one-dimensional case was considered in~\cite{DS,Sm3}. In general, it was not obvious whether ideas and methods developed in~\cite{She2}--\cite{She4} for Darboux transformations in the continuous case will allow to obtain factorization results in the discrete case. Positive answer to this question is given in this paper.

The main results on Darboux transformations and on factorization of differential operators in the continuous case are reviewed in Section~\ref{sectcont}. Classification of elementary Darboux transformations for second-order hyperbolic operators in two variables (both in the semidiscrete case and in the entirely discrete case) is given in Section~\ref{sectelem}. Some important results on factorization of difference operators as well as on Laplace invariants and Laplace series for hyperbolic operators in the discrete case are discussed in Section~\ref{sectdicr}. The main factorization theorem for Darboux--Laplace transformations in the (semi)discrete case is proved in Sections~\ref{sectfactsd} and~\ref{sectfactd}.

\section{Continuous case}\label{sectcont}

Darboux transformations may be defined in very general settings.

\begin{definition}
\rm
Let $\mathfrak K$ be a differentially closed differential field of zero characteristic with commuting derivations $\pa_{x_1},\pa_{x_2},\dots,\pa_{x^k}$. Denote the
ring of linear partial differential operators with coefficients from $\mathfrak K$ by ${\mathfrak K}[\pa_{x_1},\pa_{x_2},\dots,\pa_{x^k}]$. Operators ${\cal L}$ and $\hat{\cal L}$ are {\it related by a Darboux transformation} if there exist ${\cal D},\hat{\mathcal D}\in {\mathfrak K}[\pa_{x_1},\pa_{x_2},\dots,\pa_{x^k}]$ such that the following operator relation is satisfied:
$$
\hat{\cal L}\circ{\cal D}=\hat{\cal D}\circ{\cal L}.
$$
\end{definition}

The main property of a Darboux transformation is that it maps the kernel of the initial operator $\cal L$ into the kernel of the transformed operator
$\hat{\cal L}$. Therefore, use of Darboux transformations, in principle, allows to generate new solutions of linear partial differential equations starting
from the known ones. Although these ideas date back to Laplace, Darboux or even Euler, Darboux transformations play an important role in modern theory of integrable systems.

Operator relation~(\ref{dlt}) is equivalent to a non-linear system of partial differential equations for coefficients of the transformation operator $\cal D$ and its companion $\hat{\cal D}$. Hence, the classification problem for Darboux transformations, in general, leads to the necessity to solve such system. One can easily check by comparing the number of equations with the number of free coefficients that in the one-dimensional case ($k=1$) theory of Darboux transformations in the above settings is empty inane for arbitrary generic pair $({\cal L},{\cal D})$ the operator ${\cal D}$ is a transformation operator for ${\cal L}$. Therefore in this case the relation~(\ref{dlt}) is usually replaced by the relation $\hat{\cal L}\circ{\cal D}=\cal D\circ{\cal L}$ (transformation operator coincides with its companion), and this leads to rather meaningful theory.

It is well-known that in the one-dimensional case any linear differential operator ${\cal D}$ is factorizable into a product of first-order factors: one may choose arbitrary function $\phi\in\Ker{\cal D}$ and factor out a first-order operator: ${\cal D}={\cal D}'{\cal D}_1$, where ${\cal D}_1:=\pa_x-(\ln\phi)'_x$. Moreover, such factorizations are in one-to-one correspondence with total flags in linear space $\Ker{\cal D}$, and
\begin{equation}
\label{wronskform}
{\cal D}\psi=\frac{W(\psi,\phi_1,\phi_2,\dots,\phi_d)}{W(\phi_1,\phi_2,\dots,\phi_d)},
\end{equation}
where $W(f_1,f_2,\dots,f_d)$ is a wronskian of a family of functions $f_1,f_2,\dots,f_d$ and
$\{\phi_1,\phi_2,\dots,\phi_d\}$ is an arbitrary basis in $\Ker{\cal D}$. Remarkable thing is that not only differential operators are factorizable, but so do all Darboux transformations.
\begin{theorem}\label{th1dim}\cite{AMSh}
For an arbitrary Darboux transformation~(\ref{dt}) there exist factorizations of transformation operator and its companion into first-order factors,
$$
{\cal D}={\cal D}_d {\cal D}_{d-1}\dots{\cal D}_1,\quad\hat{\cal D}=\hat{\cal D}_d \hat{\cal D}_{d-1}\dots\hat{\cal D}_1,
$$
and a sequence of operators ${\cal L}_0,{\cal L}_1,\dots,{\cal L}_d$ where ${\cal L}_0:={\cal L}$ and ${\cal L}_d:=\hat{\cal L}$ such that
${\cal L}_{j+1}{\cal D}_j={\cal D}_j{\cal L}_j$ for all $j=0,1,\dots,d$.
\end{theorem}

In the two-dimensional case the theory of Darboux transformations had been developed by Darboux in connection with his work on differential geometry of
two-dimensional hypersurfaces, see~\cite{Da}. Let $\cal L$ and $\hat{\cal L}$ be linear hyperbolic differential operators,
$$
{\cal L}=\pa_x\pa_y+a\pa_x+b\pa_y+c,\quad\hat{\cal L}=\pa_x\pa_y+\hat a\pa_x+\hat b\pa_y+\hat c,
$$
where the coefficients are functions of two independent variables $x$ and $y$, and  $\cal D$, $\hat{\cal D}$ be first order operators

$$
{\cal D}=\alpha\pa_x+\beta\pa_y+\gamma,\quad\hat{\cal D}=\hat\alpha\pa_x+\hat\beta\pa_y+\hat\gamma.
$$

Darboux was interested in classification of such transformations and he had conjectured that there are only two types of them: classical {\it Laplace
transformations} and transformations defined by {\it wronskian formulas}. Although some partial classification results where known before, complete proof of this conjecture has been published only a few years ago, see~\cite{She1,She2}.

\begin{theorem}\label{elemdlt}{\rm\cite{She1}}
Differential first-order operator ${\cal D}=\pa_x+\gamma$ is a transformation operator for ${\cal L}=\pa_x\pa_y+a\pa_x+b\pa_y+c$ if and only if the function $\gamma$ is defined by one of the following conditions:

i) $\gamma=b$;

ii) $\gamma=-(\ln\phi)'_x$ where $\phi$ is an arbitrary function from $\Ker{\cal L}$.\\
In each of these cases suitable choice of the coefficients $\hat a$, $\hat b$, $\hat c$ and $\hat\gamma$
uniquely defines the operators $\hat{\cal L}$ and $\hat{\cal D}$. Similarly, the operator ${\cal D}=\pa_y+\gamma$ defines a Darboux transformation for ${\cal L}$ if and only if $\gamma=a$ or $\gamma=-(\ln\phi)'_y$ for some $\phi\in\Ker{\cal L}$.
\end{theorem}
\begin{remark}
\rm
Although, probably, this theorem has not been published before~2012, the classification result itself had been known to some specialists as a ``mathematical folklore'' long before. In particular, it had been communicated to the author by A.\,B.\,Shabat already in~1996.
\end{remark}
\begin{remark}
\rm
Zero-order Darboux transformations are nothing but gauge transformations
\begin{equation}
\label{gauge}
{\cal L}\mapsto\hat{\cal L}=\frac{1}{\omega}\circ{\cal L}\circ\omega,
\end{equation}
where ${\cal D}=\hat{\cal D}=\omega\in\mathfrak K$.
\end{remark}

Darboux transformations defined in the theorem~\ref{elemdlt} are called {\it elementary}. Transformations of the first type are nothing but classical {\it Laplace transformations}. Examples of non-elementary Darboux transformations defined by {\it wronskian formulas} were already known to Darboux. Therefore in the two-dimensional case transformations of a second-order hyperbolic operator defined by formula~(\ref{dlt}) are sometimes called {\it Darboux--Laplace transformations}. The following theorem provides complete description of first order Darboux transformations\footnote{In~\cite{She2} such transformations are called transformations of {\it total order} two or transformations of {\it bi-degree} two.} for hyperbolic second order operators.

\begin{theorem}{\rm\cite{She2}}
Let $\phi_1,\phi_2$ be two arbitrary functions from $\Ker{\cal L}$, where ${\cal L}=\pa_x\pa_y+a\pa_x+b\pa_y+c$. Then the operator ${\cal D}$ defined by formula
\begin{eqnarray}
\label{contwronsk}
{\cal D}\psi=\det\left(
\begin{array}{ccc}
\psi &\psi_x &\psi_y\\
\phi_1 &\phi_{1,x} &\phi_{1,y}\\
\phi_2 &\phi_{2,x} &\phi_{2,y}
\end{array}
\right)
\end{eqnarray}
is a Darboux transformation operator for ${\cal L}$. Any non-elementary first-order transformation operator ${\cal D}=\alpha\pa_x+\beta\pa_y+\gamma$ up to gauge equivalence has the form~(\ref{contwronsk}) for some $\phi_1,\phi_2\in\Ker{\cal L}$.
\end{theorem}

Darboux transformations defined by formula~(\ref{contwronsk}) or are called {\it wronskian}. Wronskian Darboux transformations of order higher than one are defined similarly. They generalize the formula~(\ref{wronskform}) for two-dimensional case. Wronskian transformations of arbitrary order were widely used by Darboux~\cite{Da}.

It is easy to check that in the one-dimensional case any Darboux transformation is a composition of a gauge transformation and $d$ elementary transformations defined by operators of the form ${\cal D}=\pa_x-(\ln\phi)'_x$, where $d$ is the order of the transformation operator and $\phi$ belongs to the kernel of corresponding operator ${\cal L}$. Hence, any Darboux transformation can be factorized in this case. Therefore, a natural question arises: whether the same thing happens in the two-dimensional case, i.e. are all Darboux transformations factorizable into products of some elementary building blocks? Obviously, the answer is no, if one approaches to the problem of factorization exactly in the same way as in one-dimensional case: Darboux transformation defined by wronskian formula~(\ref{contwronsk}) is not a product of elementary Darboux--Laplace transformations provided by theorem~\ref{elemdlt} just because it has order one. Although there are no exact factorization results in the two-dimensional case, on the level of equivalence classes each Darboux transformation is factorizable~\cite{She3,She4}.
\begin{definition}\cite{She2,She4}\label{defequiv}
\rm
Darboux--Laplace transformations ${\cal L}\mapsto\hat{\cal L}$ defined by operators ${\cal D}_1$ and ${\cal D}_2$ are called {\it equivalent}, if there exist
${\cal A}\in\mathfrak K [\pa_x,\pa_y]$ such that
$$
{\cal D}_2={\cal D}_1+{\cal A}{\cal L},\quad\hat{\cal D}_2=\hat{\cal D}_1+\hat{\cal L}{\cal A},
$$
where $\hat{\cal D}_1$ and $\hat{\cal D}_2$ are companion operators.
\end{definition}

Consider a hyperbolic operator ${\cal L}=\pa_x\pa_y+a\pa_x+b\pa_y+c$. Functions $k=c-ab-a_x$ and $h=c-ab-b_y$ are called the {\it Laplace invariants} of ${\cal L}$. They are invariant with respect to gauge transformations~(\ref{gauge}). It it easy to check that operator of that kind if factorizable into a pair of first-order factors if and only if at least one of its Laplace invariants vanishes: $k=0$ or $h=0$. Hence, the structure of rings ${\mathfrak K}[\pa_x]$ and
${\mathfrak K}[\pa_x,\pa_y]$ is completely different.

According to theorem~\ref{elemdlt} elementary Laplace transformations defined by operators ${\cal D}_x:=\pa_x+b$ and ${\cal D}_y:=\pa_y+a$ can be applied to any hyperbolic differential operator. One can easily verify that they are {\it invertible} if and only if $h\ne 0$ and $k\ne 0$ respectively. In this case inverse transformations
are defined by
$$
{\cal D}'_x:=-\frac{1}{h}(\pa_y+a)\quad\hbox{and}\quad{\cal D}'_y:=-\frac{1}{k}(\pa_x+b).
$$
Due to the fact that Laplace transformations defined by ${\cal D}_x$ and ${\cal D}_y$ are almost inverses of each other, operators obtained from ${\cal L}$ by a consequent application of these transformations are said to form the {\it Laplace series} of operator ${\cal L}$. These transformations act as shifts in opposite directions along the Laplace series. If one of the Laplace series vanishes at some step, then the series terminates in a factorizable operator.

\begin{theorem}\label{thfact}{\rm\cite{She3,She4}}
If the Laplace series of a hyperbolic second-order operator ${\cal L}$ is infinite, then every Darboux--Laplace transformation ${\cal L}\mapsto\hat{\cal L}$ is equivalent to a composition of gauge transformations and elementary transformations.
\end{theorem}

\section{Elementary Darboux--Laplace transformations}\label{sectelem}
\subsection{Semidiscrete case}

Consider a hyperbolic differential-difference operator
\begin{equation}
\label{sdhyp}
{\cal L}=\pa_x T+a_n\pa_x+b_n T+c_n,
\end{equation}
where $a_n$, $b_n$ and $c_n$ are functions depending on discrete variable $n\in\mathbb Z$ and on continuous variable $x\in\mathbb R$
and where $T$ is a shift operator: $T\psi_n (x)=\psi_{n+1}(x)$. Operators ${\cal L}$ and $\hat{\cal L}$ are {\it related by a Darboux--Laplace transformation} if there exist differential-difference operators ${\cal D}$ and $\hat{\mathcal D}$ such that relation~(\ref{dlt}) is satisfied.

In general, we are interested in developing theory of Darboux--Laplace transformations in the discrete case. As a first step in this direction, we need to classify elementary transformations. This is done by a straightforward approach based on careful analysis of the system of differential-difference equations equivalent to operator relation~(\ref{dlt}). Let
$$
{\cal D}=\alpha_n\pa_x+\beta_n T+\gamma_n,\quad\hat{\cal D}=\hat\alpha_n\pa_x+\hat\beta_n T+\hat\gamma_n,\quad\hat{\cal L}=\pa_x T+\hat a_n\pa_x+\hat b_n T+\hat c_n;
$$
then equation~(\ref{dlt}) is equivalent to the following system:
\begin{numcases}{}
\alpha_{n+1}-\hat\alpha_n=0
\label{sdeqns1}\\
\beta_{n+1}-\hat\beta_n=0
\label{sdeqns2}\\
\hat a_n\alpha_n-\hat\alpha_n a_n=0
\label{sdeqns3}\\
\hat b_n\beta_{n+1}+\beta_{n+1,x}-\hat\beta_n b_{n+1}=0
\label{sdeqns4}\\
\alpha_{n+1,x}+\hat b_n\alpha_{n+1}+\hat a_n\beta_n+\gamma_{n+1}-\hat\gamma_n-\hat\beta_n a_{n+1}-\hat\alpha_n b_n=0
\label{sdeqns5}\\
\hat a_n\alpha_{n,x}+\hat a_n\gamma_n+\hat c_n\alpha_n-\hat\alpha_n a_{n,x}-\hat\gamma_n a_n-\hat\alpha_n c=0
\label{sdeqns6}\\
\hat a_n\beta_{n,x}+\hat c_n\beta_n+\gamma_{n+1,x}+\hat b_n\gamma_{n+1}-\hat\alpha_n b_{n,x}-\hat\gamma_n b_n-\hat\beta_n c_{n+1}=0
\label{sdeqns7}\\
\hat a_n\gamma_{n,x}+\hat c_n\gamma_n-\hat\alpha_n c_{n,x}-\hat\gamma_n c_n=0
\label{sdeqns8}
\end{numcases}
Here and further $x$ in subscripts stands for the derivative with respect to $x$.
\begin{theorem}\label{thsd}
Differential operator ${\cal D}=\pa_x+\gamma_n$ is a transformation operator for~(\ref{sdhyp}) if and only if the function $\gamma_n$ is defined by one of the following conditions:

i) $\gamma_n=b_{n-1}$;

ii) $\gamma_n=-(\ln\phi_n)_x$ where $\phi$ is an arbitrary function from $\Ker{\cal L}$.\\
In each of these cases suitable choice of the coefficients $\hat a_n$, $\hat b_n$, $\hat c_n$ and $\hat\gamma_n$
uniquely defines the operators $\hat{\cal L}$ and $\hat{\cal D}$.
\end{theorem}

Proof of this theorem is based on a number of lemmas. Note that $\alpha\equiv 1$ and $\beta\equiv 0$ for transformations of the form ${\cal D}=\pa_x+\gamma_n$ that we consider now. Hence, equations~(\ref{sdeqns1}),~(\ref{sdeqns2}) and~(\ref{sdeqns4}) are automatically satisfied and equation~(\ref{sdeqns3}) yields $\hat a_n=a_n$.
\begin{definition}
\rm
Functions $k_n:=\frac{c_n}{a_n}-(\ln a_n)_x-b_n$ and  $h_n:=\frac{c_n}{a_n}-b_{n-1}$ are called {\it the Laplace invariants} of hyperbolic operator~(\ref{sdhyp}).
\end{definition}
\begin{lemma}\label{lemlapl}
If $\gamma_n=b_{n-1}$, then the following formulas define the solution of system~(\ref{sdeqns1})--(\ref{sdeqns8}):
\begin{eqnarray}
\label{aab}
\left\{
\begin{array}{l}
\hat a_n=\hat a_n\\
\hat b_n=\hat\gamma_n=b_{n-1}-(\ln (a_n h_n))_x\\
\hat c_n=c_n+a_{n,x}-a_n (\ln (a_n h_n))_x
\end{array}
\right..
\end{eqnarray}
\end{lemma}
{\bf Proof}.

If $\gamma_n=b_{n-1}$, then equation~(\ref{sdeqns4}) yields $\hat\gamma_n=\hat b_n$. Eliminate
$$
\hat c_n=c_n+a_{n,x}+a_n (\hat b_n-b_{n-1})
$$
from~(\ref{sdeqns6}) and substitute this expression into~(\ref{sdeqns8}):
$$
(a_n b_{n-1}-c_n)_x+(\hat b_n-b_{n-1})(a_n b_{n-1}-c_n)=0.
$$
This leads to formulas~(\ref{aab}). $\Box$

\begin{lemma}\label{lemh}
If $a_n\ne 0$ and equations~(\ref{sdeqns3}),~(\ref{sdeqns5}) are satisfied for the transformation of the form ${\cal D}=\pa_x+\gamma_n$,
then~(\ref{sdeqns6}) can be rewritten as follows:
$$
u_n-u_{n+1}=\hat k_n-h_n,
$$
where $u_n:=b_{n-1}-\gamma_n$.
\end{lemma}
{\bf Proof}.

Eliminate $\hat\gamma_n$ from~(\ref{sdeqns5}),
$$
\hat\gamma_n=\hat b_n-u_{n+1},
$$
substitute it into~(\ref{sdeqns6}) and use the relation $\hat a_n=a_n$. $\Box$

\begin{lemma}\label{lemk}
If $a_n\ne 0$, $u_n\ne 0$ and equations~(\ref{sdeqns3}),~(\ref{sdeqns5})--(\ref{sdeqns7}) are satisfied for the transformation of the form ${\cal D}=\pa_x+\gamma_n$,
then~(\ref{sdeqns8}) can be rewritten as follows:
\begin{equation}
\label{aad}
v_{n,x}=(v_n-a_n)k_n+v_n\frac{a_{n,x}}{a_n}+\frac{u_{n+1}v_n^2}{a_n}-u_{n+1}v_n,
\end{equation}
where $v_n:=-\frac{a_n\hat k_n}{u_{n+1}}$.
\end{lemma}
{\bf Proof}.

Equations~(\ref{sdeqns5}),~(\ref{sdeqns7}) yield
\begin{equation}
\label{aac}
\hat b_n=b_n-(\ln u_{n+1})_x,\quad\hat\gamma_n=\gamma_{n+1}-(\ln u_{n+1})_x.
\end{equation}
Hence rewrite~(\ref{sdeqns8}) as follows:
$$
a_n\gamma_{n,x}+\hat c_n\gamma_n-c_{n,x}-c_n\gamma_{n+1}+c_n\frac{u_{n+1,x}}{u_{n+1}}=0.
$$
Substitute $\gamma_n=b_{n-1}-u_n$, $\hat c_n=a_n\hat k_n+a_{n,x}+a_n\hat b_n$ and use equations~(\ref{aac}) to eliminate $\hat b_n$:
$$
\hat k_n k_n+u_{n+1}k_n-\hat k_{n,x}-\hat k_n^2+\hat k_n\frac{u_{n+1,x}}{u_{n+1}}-\hat k_n u_{n+1}=0.
$$
Finally, introduce new variable $v_n:=-\frac{a_n\hat k_n}{u_{n+1}}$ to obtain~(\ref{aad}). $\Box$
\begin{lemma}
Compatibility condition for the system
\begin{eqnarray}
\label{aae}
\left\{
\begin{array}{l}
(\pa_x+b_{n-1})\phi_n=u_n\phi_n\\
(T+a_n)\phi_n=v_n\phi_n
\end{array}
\right.
\end{eqnarray}
has the form
\begin{equation}
\label{aaf}
v_{n,x}=(v_n-a_n)(u_{n+1}-u_n)+v_n(b_{n-1}-b_n)+a_n(h_n-k_n).
\end{equation}
\end{lemma}
{\bf Proof}.

System~(\ref{aae}) can be rewritten as follows:
\begin{eqnarray}
\nonumber
\left\{
\begin{array}{l}
\phi_{n,x}=(u_n-b_{n-1})\phi_n\\
\phi_{n+1}=(v_n-a_n)\phi_n
\end{array}
\right..
\end{eqnarray}
``Cross-differentiation'' $T(\phi_{n,x})=\pa_x(\phi_{n+1})$ leads to equation~(\ref{aaf}). $\Box$
\begin{lemma}\label{lemcomp}
If system~(\ref{aae}) is compatible, then $\phi\in\Ker{\mathcal L}$.
\end{lemma}
{\bf Proof of theorem~\ref{thsd}}.

Consider the function $u_n=b_{n-1}-\gamma_n$. If $u_n=0$, then apply lemma~\ref{lemlapl} to obtain transformation of the first type. If $u_n\ne 0$ and $a_n\ne 0$, then according to lemma~\ref{lemh} and to lemma~\ref{lemk} system~(\ref{sdeqns1})--(\ref{sdeqns8}) can be reduced to the following one:
\begin{eqnarray}
\nonumber
\left\{
\begin{array}{l}
u_n-u_{n+1}=\frac{u_{n+1}v_n}{a_n}-h_n\\
v_{n,x}=(v_n-a_n)k_n+v_n\frac{a_{n,x}}{a_n}+\frac{u_{n+1}v_n^2}{a_n}-u_{n+1}v_n
\end{array}
\right.,
\end{eqnarray}
where $v_n=-\frac{a_n\hat k_n}{u_{n+1}}$. Straightforward calculation shows that this system yields compatibility condition~(\ref{aaf}). Hence system~(\ref{aae}) is compatible, and $u_n=(\ln\phi_n)_x+b_{n-1}$, where $\phi\in\Ker{\mathcal L}$ (see lemma~\ref{lemcomp}). Now equations~(\ref{sdeqns1})--(\ref{sdeqns8}) allow to express all coefficients in terms of $\phi$:
\begin{eqnarray}
\label{aag}
\left\{
\begin{array}{l}
\gamma_n=-(\ln\phi_n)_x\\
\hat a_n=a_n\\
\hat b_n=b_n-(\ln u_{n+1})_x\\
\hat c_n=c_n+a_{n,x}+a_n(\gamma_{n+1}-\gamma_n)-a_n(\ln u_{n+1})_x\\
\hat\gamma_n=\gamma_{n+1}-(\ln u_{n+1})_x
\end{array}
\right..
\end{eqnarray}

If $u_n\ne 0$ and $a_n=0$, then system~(\ref{sdeqns1})--(\ref{sdeqns8}) yields the equations
\begin{eqnarray}
\label{aah}
\left\{
\begin{array}{l}
\hat a_n=0\\
\hat b_n=b_n+\gamma_n-\gamma_{n+1}-(\ln c_n)_x\\
\hat c_n=c_n\\
\hat\gamma_n=\gamma_n-(\ln c_n)_x
\end{array}
\right.,
\end{eqnarray}
where $\gamma_n$ has to satisfy the following differential--difference equation:
\begin{equation}
\label{aak}
\left(\ln\frac{\gamma_{n+1}-b_n}{c_n}\right)_x=\gamma_{n+1}-\gamma_n.
\end{equation}
It is not difficult to prove that any solution of this equation has the form $\gamma_n=-(\ln\phi_n)_x$, where $\phi\in\Ker{\cal L}$.

Conversly, if $a_n\ne 0$, then for arbitrary function $\phi\in\Ker{\mathcal L}$ define $u_n:=b_{n-1}+(\ln\phi_n)_x$ and coefficients of operators $\hat{\mathcal L}$, ${\mathcal D}$ and $\hat{\mathcal D}$ using formulas~(\ref{aag}). It is easy to check that equations~(\ref{sdeqns1})--(\ref{sdeqns8}) are satisfied in this case. If $a_n\ne 0$, then for $\gamma_n=-(\ln\phi_n)_x$ define the coefficients using formulas~(\ref{aah}) and verify that equation~(\ref{aak}) is satisfied. $\Box$

In the semidiscrete case variables $x$ and $n$ enter equations not symmetrically and hence one has to consider Darboux transformations in the discrete variable separately.

\begin{theorem}\label{thsd1}
Difference operator ${\cal D}=T+\gamma_n$ is a transformation operator for ${\cal L}=\pa_x T+a_n\pa_x+b_n T+c_n$ if and only if the function $\gamma_n$ is defined by one of the following conditions:

i) $\gamma_n=a_n$;

ii) $\gamma_n=-\frac{\phi_{n+1}}{\phi_n}$ where $\phi$ is an arbitrary function from $\Ker{\cal L}$.\\
In each of these cases suitable choice of the coefficients $\hat a_n$, $\hat b_n$, $\hat c_n$ and $\hat\gamma_n$
uniquely defines the operators $\hat{\cal L}$ and $\hat{\cal D}$.
\end{theorem}

{\bf Proof}.

Exactly as in the proof of theorem~\ref{thsd} consider the function $u_n:=a_n-\gamma_n$ and examine equations~(\ref{sdeqns1})--(\ref{sdeqns8}) for
$\alpha_n=0$, $\beta_n=1$. If $u_n=0$, then $\gamma_n=a_n$ defines the solution of this system, and coefficients are as follows:
\begin{eqnarray}
\nonumber
\left\{
\begin{array}{l}
\hat a_n=\hat\gamma_n=\frac{a_{n+1}k_{n+1}}{k_n}\\
\hat b_n=b_{n+1}\\
\hat c_n=a_{n+1}k_{n+1}\left(1+\frac{b_n}{k_n}\right)
\end{array}
\right..
\end{eqnarray}

If $u_n\ne 0$ and $a_n=0$, then equations~(\ref{sdeqns1})--(\ref{sdeqns8}) yield the following differential-difference equation for the function $\gamma$:
$$
\frac{\gamma_{n+1}}{\gamma_n}+\gamma_{n+1,x}+(b_{n+1}-b_n)\gamma_{n+1}-c_{n+1}=0.
$$
It is not difficult to prove that for any of its solutions there exist $\phi\in\Ker{\mathcal L}$ such that $\gamma_n=-\frac{\phi_{n+1}}{\phi_n}$. The remaining coefficients are uniquely defined in this case:
\begin{eqnarray}
\nonumber
\left\{
\begin{array}{l}
\hat a_n=0\\
\hat b_n=b_{b+1}\\
\hat c_n=\frac{\gamma_{n+1}}{\gamma_n}c_n\\
\hat\gamma_n=\gamma_{n+1}
\end{array}
\right..
\end{eqnarray}

In the generic case $u_n\ne 0$, $a_n\ne 0$ equations~(\ref{sdeqns1})--(\ref{sdeqns8}) can be reduced to the following ones:
\begin{eqnarray}
\nonumber
\left\{
\begin{array}{l}
\frac{u_{n,x}}{u_n}+\frac{a_n k_n}{u_n}-\frac{a_n\hat h_n}{u_n}+\hat h_n+b_n-\frac{c_n}{a_n}=0\\
v_{n+1}-v_n+\hat h_n-h_n=0
\end{array}
\right.,
\end{eqnarray}
where $v_n:=-\frac{a_{n-1}\hat h_{n-1}}{u_{n-1}}$. These equations yield the equation
$$
u_{n,x}+a_n k_n-a_n h_n+a_n v_{n+1}-a_n v_n+h_n u_n+v_{n+1}u_n-u_n v_n+b_n u_n-\frac{c_n}{a_n}u_n=0,
$$
that is equivalent to compatibility conditions of the system
\begin{eqnarray}
\nonumber
\left\{
\begin{array}{l}
(\pa_x+b_{n-1})\phi_n=v_n\phi_n\\
(T+a_n)\phi_n=u_n\phi_n
\end{array}
\right..
\end{eqnarray}
Compatibility of this system implies that $\phi\in\Ker{\mathcal L}$. Hence, $u_n=a_n+\frac{\phi_{n+1}}{\phi_n}$, and the coefficients are as follows:
\begin{eqnarray}
\nonumber
\left\{
\begin{array}{l}
\gamma=-\frac{\phi_{n+1}}{\phi_n}\\
\hat a_n=\frac{u_{n+1}}{u_n}a_n\\
\hat b_n=b_{n+1}\\
\hat c_n=\frac{u_{n+1}}{u_n}\left(c_n-a_n\left(\ln\frac{\phi_{n+1}}{\phi_n}\right)\right)\\
\hat\gamma_n=\frac{u_{n+1}}{u_n}\gamma_n
\end{array}
\right..
\end{eqnarray}
This gives Darboux transformations of the second type. $\Box$

\subsection{Purely discrete case}

In the purely discrete case consider a hyperbolic difference operator
\begin{equation}
\label{dhyp}
{\cal L}=T_1 T_2+a_{n,m}T_1+b_{n,m}T_2+c_{n,m},
\end{equation}
where $a_{n,m}$, $b_{n,m}$ and $c_{n,m}$ are functions depending on discrete variables $n,m\in\mathbb Z$ and where $T_1$, $T_2$ are shift operators in variables $n$ and $m$ respectively. Similarly to the previous cases Darboux transformations are defined by operator relation~(\ref{dlt}), where
$$
{\cal D}=\alpha_{n,m}T_1+\beta_{n,m} T_2+\gamma_{n,m},\quad\hat{\cal D}=\hat\alpha_{n,m}T_1+\hat\beta_{n,m} T_2+\hat\gamma_{n,m},\quad\hat{\cal L}=T_1 T_2+\hat a_{n,m}T_1+\hat b_{n,m} T_2+\hat c_{n,m}.
$$
This operator relation is equivalent to the following system of equations:
\begin{eqnarray}
\label{deqns}
\left\{
\begin{array}{l}
\hat\alpha_{n,m}-\alpha_{n+1,m+1}=0\\
\hat\beta_{n,m}-\beta_{n+1,m+1}=0\\
\hat a_{n,m}\alpha_{n+1,m}-\hat\alpha_{n,m}a_{n+1,m}=0\\
\hat b_{n,m}\beta_{n,m+1}-\hat\beta_{n,m}b_{n,m+1}=0\\
\hat b_{n,m}\alpha_{n,m+1}+\hat a_{n,m}\beta_{n+1,m}+\gamma_{n+1,m+1}-\hat\gamma_{n,m}-\hat\beta_{n,m}a_{n,m+1}-\hat\alpha_{n,m}b_{n+1,m}=0\\
\hat c_{n,m}\alpha_{n,m}+\hat a_{n,m}\gamma_{n+1,m}-\hat\gamma_{n,m}a_{n,m}-\hat\alpha_{n,m}c_{n+1,m}=0\\
\hat c_{n,m}\beta_{n,m}+\hat b_{n,m}\gamma_{n,m+1}-\hat\gamma_{n,m}b_{n,m}-\hat\beta_{n,m}c_{n,m+1}=0\\
\hat c_{n,m}\gamma_{n,m}-\hat\gamma_{n,m}c_{n,m}=0
\end{array}
\right.
\end{eqnarray}

Exactly in the same way as for differential-difference operator, we are going to classify elementary Darboux--Laplace transformations in the entirely discrete case. {\it Laplace invariants} are defined for operators of this kind as follows:
$$
h_{n,m}=\frac{c_{n,m}}{a_{n,m} b_{n,m-1}}-1,\quad k_{n,m}=\frac{c_{n,m}}{a_{n,m} b_{n-1,m}}-1.
$$
\begin{theorem}\label{thd}
Difference operator ${\cal D}=T_1+\gamma_{n,m}$ is a transformation operator for~(\ref{dhyp}) if and only if the function $\gamma$ is defined by one of the following conditions:

i) $\gamma_{n,m}=b_{n,m-1}$;

ii) $\gamma_{n,m}=-\frac{\phi_{n+1,m}}{\phi_{n,m}}$ where $\phi$ is an arbitrary function from $\Ker{\cal L}$.\\
In each of these cases suitable choice of the coefficients $\hat a_{n,m}$, $\hat b_{n,m}$, $\hat c_{n,m}$ and $\hat\gamma_{n,m}$
uniquely defines the operators $\hat{\cal L}$ and $\hat{\cal D}$.

Difference operator ${\cal D}=T_2+\gamma_{n,m}$ is a transformation operator for~(\ref{dhyp}) if and only if the function $\gamma$ is defined by one of the following conditions:

i) $\gamma_{n,m}=b_{n-1,m}$;

ii) $\gamma_{n,m}=-\frac{\phi_{n,m+1}}{\phi_{n,m}}$ where $\phi$ is an arbitrary function from $\Ker{\cal L}$.\\
In each of these cases suitable choice of the coefficients $\hat a_{n,m}$, $\hat b_{n,m}$, $\hat c_{n,m}$ and $\hat\gamma_{n,m}$
uniquely defines the operators $\hat{\cal L}$ and $\hat{\cal D}$.
\end{theorem}
{\bf Proof}.

Discrete variables $n$ and $m$ enter all equations~(\ref{deqns}) symmetrically and therefore it is sufficient to prove only the first part of this theorem ($\alpha_{n,m}=1$ and $\beta_{n,m}=0$ in this case). Denote $u_{n,m}:=b_{n,m-1}-\gamma_{n,m}$. If $u_{n,m}=0$, then the formulas
\begin{eqnarray}
\nonumber
\left\{
\begin{array}{l}
\gamma_{n,m}=b_{n,m-1}\\
\hat a_{n,m}=a_{n+1,m}\\
\hat b_{n,m}=\hat\gamma_{n,m}=b_{n+1,m-1}\frac{a_{n+1,m} h_{n+1,m}}{a_{n,m} h_{n,m}}\\
\hat c_{n,m}=\frac{a_{n+1,m} b_{n+1,m-1} h_{n+1,m}}{a_{n,m} b_{n,m-1} h_{n,m}} c_{n,m}
\end{array}
\right.
\end{eqnarray}
define a solution of equations~(\ref{deqns}).

If $u_{n,m}\ne 0$, then introduce new variable $v_{n,m}:=-\frac{a_{n-1,m} b_{n-1,m}\hat k_{n-1,m}}{u_{n-1,m+1}}$. Equations~(\ref{deqns}) yield the relation
$$
a_{n,m}u_{n,m}-a_{n-1,m}u_{n,m+1}+a_{n-1,m}b_{n,m}-a_{n,m}b_{n,m-1}+b_{n,m-1}v_{n+1,m}-u_{n,m}v_{n+1,m}-b_{n,m}v_{n,m}+u_{n,m+1}v_{n,m}=0,
$$
which is equivalent to the compatibility condition for the system
\begin{eqnarray}
\nonumber
\left\{
\begin{array}{l}
(T_1+b_{n,m-1})\phi_{n,m}=u_{n,m}\phi_{n,m}\\
(T_2+a_{n-1,m})\phi_{n,m}=v_{n,m}\phi_{n,m}
\end{array}
\right..
\end{eqnarray}
Express now $u_{n,m}$ in terms of a solution, $u_{n,m}=b_{n,m-1}+\frac{\phi_{n+1,m}}{\phi_{n,m}}$, and determine all coefficients:
\begin{eqnarray}
\label{aam}
\left\{
\begin{array}{l}
\gamma_{n,m}=-\frac{\phi_{n+1,m}}{\phi_{n,m}}\\
\hat\gamma_{n,m}=\frac{u_{n+1,m+1}}{u_{n,m+1}}b_{n,m}-u_{n+1,m+1}\\
\hat a_{n,m}=a_{n+1,m}\\
\hat b_{n,m}=\frac{u_{n+1,m+1}}{u_{n,m+1}}b_{n,m}\\
\hat c_{n,m}=\frac{\hat\gamma_{n,m}}{\gamma_{n,m}}c_{n,m}
\end{array}
\right..
\end{eqnarray}
Cases $a_{n,m}=0$ and $b_{n,m}=0$ has to be considered separately. $\Box$

\section{Factorization of Darboux--Laplace transformations}

\subsection{Factorization and Laplace series for difference operators}\label{sectdicr}

In the one-dimensional case any linear difference operator is factorizable into a product of first-order factors and every Darboux transformation is also factorizable. This can be proved absolutely in the same way as theorem~\ref{th1dim}. In this case, the first-order operator that has to be factored out from the initial operator ${\cal D}$ has the form $T-\frac{\phi_{n+1}}{\phi_n}$, where $\phi_n\in\Ker{\cal D}$. Wronskian formula~(\ref{wronskform}) has to be replaced by the following discrete wronskian or {\it casoratian} formula:
\begin{equation}
\label{casfor}
{\cal D}\psi=\frac{C(\psi_n,\phi_n^1,\phi_n^2,\dots,\phi_n^d)}{C(\phi_n^1,\phi_n^2,\dots,\phi_n^d)},
\end{equation}
where $\{\phi_n^1,\phi_n^2,\dots,\phi_n^d\}$ forms a basis in $\Ker{\cal D}$ and
\begin{eqnarray}
\nonumber
C(f_n^1,f_n^2,\dots,f_n^d):=\det\left(
\begin{array}{ccccc}
f_n^1 & f_{n+1}^1 & f_{n+2}^1 &\dots & f_{n+d-1}^1\\
f_n^1 & f_{n+1}^2 & f_{n+2}^2 &\dots & f_{n+d-1}^2\\
\vdots & \vdots & \vdots & \ddots & \vdots\\
f_n^d & f_{n+1}^d & f_{n+2}^d &\dots & f_{n+d-1}^d
\end{array}
\right)
\end{eqnarray}
is the {\it Casorati determinant} of a family of functions $f_n^1,f_n^2,\dots,f_n^d$.

Laplace invariants for two-dimensional hyperbolic differential-difference and entirely difference operators were defined in Section~\ref{sectelem}. Similarly to the continuous case, they are invariant with respect to gauge transformations ${\cal L}\mapsto\omega_{n+1}^{-1}{\cal L}\omega_n$ for semidiscrete operators
(and ${\cal L}\mapsto\omega_{n+1,m+1}^{-1}{\cal L}\omega_{n,m}$ for entirely discrete operators). Laplace invariants control factorizability of hyperbolic operators in the discrete case as well. Elementary transformations of the first type in different variables that were introduced in theorems~\ref{thsd},~\ref{thsd1},~\ref{thd}) are inverse to each other (see~\cite{AS} for details). This allows to define the Laplace series of a (semi)discrete hyperbolic operator exactly in the same way as in the continuous case.

Casoratian formulas~(\ref{casfor}) can easily be generalized for the two-dimensional case, cf.~(\ref{contwronsk}). Such two-dimensional casoratian formulas (or, equivalently, Darboux transformations) where used in~\cite{DT} for obtaining explicit solutions of entirely discrete Toda lattices corresponding to $A$-series Cartan matrices.

\subsection{Semidiscrete case}\label{sectfactsd}

In the semidiscrete case the theory of Darboux transformations does not differ much from its continuous analog. Darboux--Laplace transformations of hyperbolic second-order operators can be factorized only on the level of equivalence classes in the sense of definition~\ref{defequiv}. The main purpose of this section is to prove the following theorem.

\begin{theorem}\label{thfactsd}
If the Laplace series of hyperbolic second-order operator~(\ref{sdhyp}) is infinite, then every Darboux--Laplace transformation ${\cal L}\mapsto\hat{\cal L}$ is equivalent to a composition of gauge transformations and elementary transformations.
\end{theorem}

{\bf Proof}.

Proof of this theorem is very similar to the proof of theorem~\ref{thfact} given in~\cite{She3,She4}. The only difference comes from the fact that difference operators do not satisfy Leibniz product rule.

Consider Darboux--Laplace transformation $\hat{\cal L}{\cal D}=\hat{\cal D}{\cal L}$, where ${\cal D}$ and $\hat{\cal D}$ are arbitrary differential-difference operators:
$$
{\cal D}=\sum\limits_{i,j}^{}\gamma_{ij}\pa_x^i T^j,\quad\hat{\cal D}=\sum\limits_{i,j}^{}\hat\gamma_{ij}\pa_x^i T^j.
$$
One can easily verify that by proper choice of the operator ${\cal A}$ this Darboux--Laplace transformation can be made equivalent to the transformation such that the operators ${\cal D}$ and $\hat{\cal D}$ do not contain mixed terms with both $\pa_x^j$ and $T^j$. Therefore on the level of equivalence classes it is sufficient to consider only Darboux--Laplace transformations with ${\cal D}$ and $\hat{\cal D}$ of the following form:
$$
{\cal D}=\sum\limits_{i=1}^{d_1} \alpha_i\pa_x^i+\sum\limits_{j=1}^{d_2} \beta_j T^j+\gamma,\quad
\hat{\cal D}=\sum\limits_{i=1}^{d_1} \hat\alpha_i\pa_x^i+\sum\limits_{j=1}^{d_2} \hat\beta_j T^j+\hat\gamma.
$$

The next step is to reduce this Darboux--Laplace transformation to transformation with ${\cal D}'$ and $\hat{\cal D}'$ being operators only in $\pa_x$ or only in $T$. If the Laplace series is infinite (or at least is long enough) this could be done using elementary Laplace transformations of the first type (see theorems~\ref{thsd},~\ref{thsd1}). More precisely, consider Laplace transformation ${\cal L}_1 {\cal D}_1=\hat{\cal D}_1{\cal L}$, where ${\cal D}_1:=\pa_x+b_n$ and $\hat{\cal D}_1=\pa_x+\hat b_n$. It is easy to check that the product of this transformation and the initial one is equivalent to a transformation with operators ${\cal D}'$ and $\hat{\cal D}'$ having degree $d_1+1$ in powers of $\pa_x$ and degree $d_2-1$ in powers of $T$. Repeat such procedure $d_2$ times to obtain transformation operators with no powers of $T$ (although their coefficients can depend on both variables). Since the Laplace series of the initial operator is infinite, all elementary Laplace transformations are invertible.

Now one has to factorize transformations of the form
\begin{equation}
\label{aal}
{\cal D}=\sum\limits_{i=1}^d \alpha_i\pa_x^i+\gamma,\quad
\hat{\cal D}=\sum\limits_{i=1}^d \hat\alpha_i\pa_x^i+\hat\gamma.
\end{equation}
If $\Ker{\cal L}\cap\Ker{\cal D}\ne\{0\}$, then consider ${\cal D}_1:=\pa_x-\frac{\phi_{n,x}}{\phi_n}$, where $\phi_n\in\Ker{\cal L}\cap\Ker{\cal D}$. In this case operator ${\cal D}$ is divisible by ${\cal D}_1$: ${\cal D}={\cal D}'{\cal D}_1$ for some ${\cal D}'$. Using appropriate gauge transformation
${\cal L}\mapsto\tilde{\cal L}:=\omega_{n+1}^{-1}{\cal L}\omega_n$ one can obtain hyperbolic operator of the form $\tilde{\cal L}=\pa_x T+a_n\pa_x+c_n$. Therefore it is sufficient to consider only operators of this kind. In this case if $c_n\ne 0$, then equations~(\ref{aag}) yield $\hat\gamma_n=-\ln (\phi_{n+1,x})_x$ (otherwise one of the Laplace invariants vanishes, and this contradicts our assumption). Introduce operator $\hat{\cal D}_1:=\pa_x-\ln (\phi_{n+1,x})_x$ and use relation~(\ref{dlt}):
$$
\hat{\cal D}{\cal L}(n\phi_n)=\hat{\cal L}{\cal D} (n\phi_n)=\hat{\cal L}\left(n{\cal D}(\phi_n)\right)=0
$$
since ${\cal D}$ is an operator only in $\pa_x$ and $\phi_n\in\Ker\hat{\cal D}$. On the other hand,
$$
\hat{\cal D}{\cal L}(n\phi_n)=\hat{\cal D}\left((n+1)\phi_{n+1,x}+n a_n\phi_{n,x}+n c_n\phi_n\right)=\hat{\cal D}\left((n+1)\phi_{n+1,x}-n\phi_{n+1,x}\right)=
\hat{\cal D}\left(\phi_{n+1,x}\right).
$$
Hence, $\phi_{n+1,x}\in\Ker\hat{\cal D}$ and therefore $\hat{\cal D}$ is divisible by $\hat{\cal D}_1$: $\hat{\cal D}=\hat{\cal D}'\hat{\cal D}_1$. This means that operator relation~(\ref{dlt}) yields
$$
\hat{\cal L}{\cal D}'{\cal D}_1=\hat{\cal D}'\hat{\cal D}_1{\cal L}=\hat{\cal D}'{\cal L}_1{\cal D}_1\quad\Longrightarrow\quad
\hat{\cal L}{\cal D}'=\hat{\cal D}'{\cal L}_1,
$$
and the initial Darboux transformation is a product of a gauge transformation, elementary transformation and transformation of the order $d-1$ defined by operators ${\cal D}'$ and $\hat{\cal D}'$, i.e. we have managed to factor out one elementary Darboux transformation and to reduce the order of the transformation by one.

Consider now the case $\Ker{\cal L}\cap\Ker{\cal D}=\{0\}$. Let ${\cal D}_x:=\pa_x+b_{n-1}$ be the elementary Laplace transformation operator and suppose that $\Ker{\cal D}\cap\Ker{\cal D}_x=\{0\}$. Then the restriction ${\cal D}_x|_{\Ker{\cal D}}$ is a non-degenerate $\Omega_n$-linear mapping, where $\Omega_n$ is the space of all functions depending only on the variable $n$ (operator ${\cal D}$ does not contain powers of $T$). One can easily verify that
\begin{equation}
\label{leib}
{\cal L}(f_n\psi_n)=f_n{\cal L}(\psi_n)+(f_{n+1}-f_n)T{\cal D}_x (\psi_n)
\end{equation}
for any $f_n\in\Omega_n$ and for arbitrary function $\psi_n=\psi_n(x)$. Using formula~(\ref{leib}) and relation~(\ref{dlt}) one may deduce that
$$
\hat{\cal D}\left(T{\cal D}_x(\psi_n)\right)=\hat{\cal D}{\cal L}(n\psi_n)-\hat{\cal D} \left(n{\cal L}\psi_n\right)=
\hat{\cal L}{\cal D}(n\psi_n)-n\hat{\cal D}{\cal L}(\psi_n)=0-n\hat{\cal L}{\cal D}(\psi_n)=0
$$
for any $\psi_n\in\Ker{\cal D}$. Since both $\Ker{\cal D}$ and $\Ker\hat{\cal D}T$ are $d$-dimensional left modules over $\Omega_n$, restriction
${\cal D}_x|_{\Ker{\cal D}}$ is an isomorphism $\Ker{\cal D}\to\Ker\hat{\cal D}T$.

Relation~(\ref{dlt}) implies that ${\cal L}(\psi_n)\in\Ker\hat{\cal D}$ for any $\psi_n\in\Ker{\cal D}$. Prove now by induction that there exists a basis $\{e_n^1,\dots,e_n^d\}$ in the $\Omega_n$-module $\Ker{\cal D}$ such that $\{u_n^1,\dots,u_n^d\}$ form a basis in $\Ker\hat{\cal D}$, where $u_n^j:={\cal L}(e_n^j)$ for all $j=1,\dots,d$. The base follows from the fact that $\Ker{\cal L}\cap\Ker{\cal D}=\{0\}$. Suppose vectors $\{e_n^1,\dots,e_n^k\}$ with required properties are already found and take any $e_n^{k+1}\in\Ker{\cal D}$. If $\{u_n^1,\dots,u_n^{k+1}\}$ are $\Omega_n$-linearly independent, then the inductive step is checked. Otherwise, denote $v_n^j:={\cal D}_x (e_n^j)$ for all $j=1,\dots,k+1$. These vectors are $\Omega_n$-linearly independent in $\Ker\hat{\cal D}T$ since
${\cal D}_x|_{\Ker{\cal D}}$ is an isomorphism of $\Omega_n$-modules. Therefore vectors $v_{n+1}^1,\dots,v_{n+1}^{k+1}$ are $\Omega_n$-linearly independent in $\Ker\hat{\cal D}$, and there exists at least one $j$ such that $v_{n+1}^j\not\in\left\langle u_n^1,\dots,u_n^{k+1}\right\rangle_{\Omega_n}$. Choose
$\tilde e_n^{k+1}:=e_n^{k+1}+ne_n^j$. Hence due to~(\ref{leib})
$$
\tilde u^{k+1}_n:={\cal L}(\tilde e^{k+1}_n)=u_n^{k+1}+nu_n^j+v_{n+1}^j
$$
is $\Omega_n$-linearly independent with $\left\langle u_n^1,\dots,u_n^k\right\rangle_{\Omega_n}$. This proves the existence of the basis $\{u_n^1,\dots,u_n^d\}$.

Consider arbitrary $\psi_n=\psi_n^1 e_1+\dots+\psi_n^d e_d\in\Ker{\cal D}$. Using formula~(\ref{leib}) one may obtain that
\begin{multline*}
{\cal L}(\psi_n)=\sum_{i=1}^d \left(\psi_n^i {\cal L}(e_n^i)+(\psi_{n+1}^i-\psi_n^i)T{\cal D}_x (e_n^i)\right)=
\sum_{i=1}^d \left(\psi_n^i u_n^i+(\psi_{n+1}^i-\psi_n^i)v_{n+1}^i\right)=\\
=\sum_{i=1}^d \left(\psi_n^i u_n^i+(\psi_{n+1}^i-\psi_n^i)\sum_{j=1}^d a_{ij} u_n^j\right)=
\sum_{j=1}^d \left(\sum_{i=1}^d a_{ij}(\psi_{n+1}^i-\psi_n^i)+\psi_n^j\right)u_n^j,
\end{multline*}
where $A=(a_{ij})$ is the matrix of linear mapping $\Ker{\cal D}\to\Ker\hat{\cal D}$ defined by identification of bases: $Au_n^k=v_{n+1}^k$. Since
${\cal D}_x|_{\Ker{\cal D}}$ is non-degenerate, $\det A\ne 0$ and the system $A(\mb x_{n+1}-\mb x_n)+\mb x_n=0$, where $\mb x_n:=(\psi_n^1,\dots,\psi_n^d)^t$, has non-trivial solution. This contradicts the fact that vectors $e_n^1,\dots,e_n^d$ are $\Omega_n$-linearly independent. Hence,
$\Ker{\cal D}\cap\Ker{\cal D}_x\ne\{0\}$.

Since ${\cal D}_x$ is a first-order operator, condition $\Ker{\cal D}\cap\Ker{\cal D}_x\ne\{0\}$ implies $\Ker{\cal D}_x\subset\Ker{\cal D}$ and therefore differential operator ${\cal D}$ is divisible by ${\cal D}_x$: there exists ${\cal D}'$ such that ${\cal D}={\cal D}'{\cal D}_x$ and $\ord{\cal D}'=d-1$. Consider arbitrary $\phi_n\in \Ker{\cal D}\cap\Ker{\cal D}_x$. The assumption $\Ker{\cal L}\cap\Ker{\cal D}=\{0\}$ and relations
$$
\hat{\cal D}_x{\cal L}(\phi_n)={\cal L}_1{\cal D}_x(\phi_n)=0,\quad\hat{\cal D}{\cal L}(\phi_n)=\hat{\cal L}{\cal D}(\phi_n)=0
$$
yield $\Ker\hat{\cal D}_x\subset\Ker\hat{\cal D}$ and therefore $\hat{\cal D}$ is divisible by $\hat{\cal D}_x$: $\hat{\cal D}=\hat{\cal D}'\hat{\cal D}_x$, where $\ord\hat{\cal D}'=d-1$. This means that the initial Darboux transformation is a product of an elementary Laplace transformation and a transformation of order $d-1$.

We have proved that in both cases $\Ker{\cal L}\cap\Ker{\cal D}\ne\{0\}$ and $\Ker{\cal L}\cap\Ker{\cal D}=\{0\}$ it is possible to factor out a first-order Darboux transformation and hence to reduce the order of the initial transformation by one, but in the first case Laplace invariants can possibly vanish at some stage of this inductive procedure. Therefore we have to prove that if the Laplace invariant $h=0$ for a hyperbolic operator ${\cal L}$ and Darboux--Laplace transformation $\hat{\cal L}{\cal D}=\hat{\cal D}{\cal L}$ is generated by operators ${\cal D}$ and $\hat{\cal D}$ of the form~(\ref{aal}), then an elementary transformation can also be factored out.

Suppose $h=0$. Hence the operator ${\cal L}$ can be factorized: ${\cal L}={\cal D}_n {\cal D}_x$, where ${\cal D}_n:=T+a_n$ and ${\cal D}_x:=\pa_x+b_{n-1}$. If $\Ker{\cal D}\cap{\cal D}_x\ne\{0\}$, then ${\cal L}(\phi_n)={\cal D}_n {\cal D}_x(\phi_n)=0$ for any $\phi_n\in\Ker{\cal D}\cap{\cal D}_x$ and hence
$\Ker{\cal L}\cap\Ker{\cal D}\ne\{0\}$. In this case the above procedure that allows to factor out one elementary transformation can be applied unless $\phi_{n,x}=0$ for $\phi_n\in\Ker{\cal L}\cap\Ker{\cal D}$. The assumption $\phi_{n,x}=0$ yields ${\cal D}_x=\pa_x$ and therefore there exists ${\cal D}'$ such that
${\cal D}={\cal D}\pa_x$. Hence, operator relation~(\ref{dlt}) is equivalent to the following one:
$$
\hat{\cal L}{\cal D}'=\hat{\cal D}(T+a_n)=\hat{\cal D}'\hat{\cal D}_1 (T+a_n)=\hat{\cal D}'{\cal L}_1,
$$
where ${\cal D}_1$ is an arbitrary first-order factor of $\hat{\cal D}$ with leading coefficient equal to $1$ and ${\cal L}_1:=\hat{\cal D}_1 (T+a_n)$. This provides the required factorization of the initial Darboux transformation and completes the proof. $\Box$

\begin{remark}
\rm
Theorem~\ref{thfactsd} also holds true for hyperbolic operators with finite Laplace series if this series is long enough to apply $d_2$ elementary Laplace transformations and hence to eliminate all terms containing shifts.
\end{remark}

\subsection{Purely discrete case}\label{sectfactd}

In the purely discrete case it is also possible to factorize Darboux--Laplace transformations.
\begin{theorem}
If the Laplace series of hyperbolic second-order operator~(\ref{dhyp}) is infinite, then every Darboux--Laplace transformation ${\cal L}\mapsto\hat{\cal L}$ is equivalent to a composition of gauge transformations and elementary transformations.
\end{theorem}

{\bf Proof}.

Proof of this theorem is almost the same as the proof of theorem~\ref{thfactsd} given in the previous subsection. We will only highlight some minor differences arising from the fact that everything is discrete now. Applying a sufficient number of Laplace transformations one can reduce generic case to the case when transformation operators contain shifts only in one direction (say, in direction $n$). If $\Ker{\cal L}\cap\Ker{\cal D}\ne\{0\}$, then one can choose arbitrary $\phi_{n,m}\in\Ker{\cal L}\cap\Ker{\cal D}$ and show that ${\cal D}$ is divisible by ${\cal D}_1:=T_1+\gamma_{n,m}$, where $\gamma_{n,m}=-\frac{\phi_{n+1,m}}{\phi_{n,m}}$. In the purely discrete case additional gauge transformations are senseless because they do not simplify the form of operator~(\ref{dhyp}). Nevertheless, using the formula
$$
{\cal L}(m\phi_{n,m})=\phi_{n+1,m+1}+b_{n,m}\phi_{n,m+1},
$$
where $\phi_{n,m}\in\Ker{\cal L}$, one can prove that $\hat{\cal D}(\phi_{n+1,m+1}+b_{n,m}\phi_{n,m+1})=0$. Comparing this to the expression for $\hat\gamma_{n,m}$ in~(\ref{aam}) one can deduce that $\Ker\hat{\cal D}_1\subset\Ker\hat{\cal D}$ and therefore $\hat{\cal D}$ is divisible by $\hat{\cal D}_1$. This allows to factor out an elementary Darboux transformation.

If $\Ker{\cal L}\cap\Ker{\cal D}=\{0\}$, then we need to factor out elementary Laplace transformation, defined by ${\cal D}_n:=T_1+b_{n,m-1}$. In this case formula~(\ref{leib}) has to be replaced by
$$
{\cal L}(f_m\psi_{n,m})=f_m{\cal L}(\psi_{n,m})+(f_{m+1}-f_m)T_2{\cal D}_n (\psi_{n,m}),
$$
where $f_m$ is a function of one variable $m$. The rest of the proof remains unchanged. $\Box$

\section{Acknowledgements}

The author is very grateful to A.~B.~Shabat for 20-year old discussions about Darboux--Laplace transformations. The work was carried out in Moscow State University
with financial support of Russian Science Foundation grant no.~16-11-10260.

\end{document}